\documentclass[twocolumn,prd,nofootinbib,aps,floats,floatfix,amsmath,amssymb,secnumarabic]{revtex4} %

\usepackage{graphicx}
\usepackage[usenames,dvipsnames]{color}
\usepackage{amsmath,amssymb}
\usepackage{slashed}
\usepackage[colorlinks,breaklinks,citecolor=blue,urlcolor=red]{hyperref}

\newcommand{\be}{\begin{equation}}
\newcommand{\ee}{\end{equation}}
\newcommand{\ba}{\begin{array}}
\newcommand{\ea}{\end{array}}
\newcommand{\bea}{\begin{eqnarray}}
\newcommand{\eea}{\end{eqnarray}}
\newcommand{\sss}{\scriptscriptstyle}
\newcommand{\ew}{e_{\sss\rm W}}

\newcommand{\DM}{{\rm DM}}

\begin{document}
\title{Dark Matter CMB Constraints and Likelihoods\\ for Poor Particle Physicists}
\author{James M.\ Cline}
\email{jcline@physics.mcgill.ca}
\affiliation{Department of Physics, McGill University,
3600 Rue University, Montr\'eal, Qu\'ebec, Canada H3A 2T8}
\author{Pat Scott}
\email{patscott@physics.mcgill.ca}
\affiliation{Department of Physics, McGill University,
3600 Rue University, Montr\'eal, Qu\'ebec, Canada H3A 2T8}
\begin{abstract}
The cosmic microwave background provides constraints on the
annihilation and decay of light dark matter at redshifts between 100 and 1000,
the strength of which depends upon the fraction of energy ending up in the
form of electrons and photons.  The resulting constraints are usually
presented for a limited selection of annihilation and decay channels.  Here we
provide constraints on the annihilation cross section and decay rate, 
at discrete values of the dark matter mass $m_\chi$, for all
the annihilation and decay channels whose secondary spectra have been computed
using PYTHIA in \href{http://arxiv.org/abs/arXiv:1012.4515}{arXiv:1012.4515}
(``PPPC4DMID: A Poor Particle Physicist Cookbook for Dark Matter
Indirect Detection''), namely $e$,
$\mu$, $\tau$,  $V\to e$, $V\to\mu$, $V\to\tau$, $u$, $d$ $s$, $c$, $b$, $t$,
$\gamma$, $g$, $W$, $Z$ and $h$.  By
interpolating in mass, these can be used to find the CMB constraints and likelihood functions 
from WMAP7 and Planck for a wide range of dark matter models,
including those with annihilation or decay into a linear combination of
different channels.
\end{abstract}
\maketitle

The temperature and polarization fluctuations of the cosmic microwave
background (CMB) are well known to be sensitive to the redshift of
recombination,  $z\approx 1100$, as this determines the surface of last
scattering.   If dark matter annihilation or decay deposits
electromagnetic energy in the primordial plasma after $z\approx 1100$,
it can delay recombination and/or contribute to reionization, leading to distortions in the CMB
\cite{Dodelson:1991bz,Scott91,Hansen:2003yj,Chen:2003gz,Padmanabhan:2005es,Mapelli:2006ej,Zhang:2007zzh,Natarajan:2009bm,
Galli:2009zc, Slatyer:2009yq,  Galli:2011rz, Cirelli:2009bb, Hutsi:2011vx,
Finkbeiner:2011dx, Natarajan:2012ry, Giesen:2012rp, Slatyer:2012yq, Frey:2013wh, Weniger:2013hja}.   This is
especially constraining for light dark matter $\chi$ with mass $m_\chi
\lesssim$ 10\,GeV (as its number density is greater than that of
heavier dark matter), and if annihilation or decay is into electrons; in that
case the current limit on the annihilation cross section is close to
the standard relic density value  $\langle\sigma v\rangle=3\times
10^{-26}$ cm$^3$/s.  Apart from the special case of monochromatic photons, constraints are generically weaker for annihilations or decays into
other particles, as those channels all involve substantial eventual yields into neutrinos and/or hadrons, neither of which efficiently transfer energy to the primordial gas (neutrinos because they are weakly-interacting, hadrons because they are strongly penetrating \cite{Chen:2003gz}).

Recent progress has been made in refining and systematizing the 
CMB bounds on dark matter annihilation and decay
\cite{Finkbeiner:2011dx,Slatyer:2012yq}.  A key  
quantity for determining the constraint on a given model is the
efficiency $f(z)$ for producing ionizing radiation, as a function
of redshift $z$. For annihilations, $f(z)$ is defined in terms of 
the electromagnetic power injected per unit volume,
\be	
   {dE\over dt\,dV} = f(z){\langle\sigma v\rangle\over m_\chi}\, 
   \Omega_{\rm DM}^2\, \rho_c^2\,c^{2}\,(1+z)^6,
\label{power}
\ee
where $\rho_c$ is the critical mass density of the universe today and $\Omega_{\rm DM}$ is the fraction in dark matter.  For perfect
efficiency, $f=1$, this is just twice the the DM mass times the annihilation rate per unit
volume. Note that no factors of two remain in (\ref{power}), as the annihilation rate itself contains a factor of one half for annihilation of identical particles \cite{Gondolo:2004sc}, which cancels the factor of two from the release of twice the DM rest mass in each annihilation event.  Here we assume Majorana DM; were the DM not its own antiparticle, (\ref{power}) would need to be divided by a further factor of two, and $\Omega_{\rm DM}$ interpreted as the total mass fraction of DM $+$ anti-DM particles.

Ref.\ \cite{Slatyer:2012yq} has provided transfer
functions $T_i(z',z,E)$\footnote{Available at
\href{http://nebel.rc.fas.harvard.edu/epsilon/}{http://nebel.rc.fas.harvard.edu/epsilon/}} that determine the contribution to $f(z)$
from particles $i=\gamma$ or $e^\pm$ pairs injected at redshift $z'$
with energy $E$.  The injected $\gamma$ or $e^\pm$ may be primary
products of dark matter annihilation or decay, or they may be
secondary particles arising from the
showering and decay of the primary ones.
 If $\phi_i=dN_i/dE$ is the electron/photon spectrum
of state $i$, normalized such that $\int dE\, E\, \phi_i$ is the fraction of 
initial dark matter mass converted into energy in that state, 
then\footnote{Our $\tilde T_i$ is related to $T_i$ of 
\cite{Slatyer:2012yq} by $\tilde T_i =
H(z) H(z')^{-1}(1+z')^n(1+z)^{-(n+1)}T_i$, where $n=2$ for annihilations and
$n=-1$ for decays.\label{fn1}}
\be
	f(z) = \sum_{i=\gamma,\,e^\pm} \int dz' \int dE\, E\, \tilde T_i(z',z,E)\, \phi_i(E).
\label{fzcurve}
\ee
The departure from perfect efficiency, $f=1$, is due to the fraction 
of initial energy that ends up in neutrinos (or possibly more exotic
invisible final states, which we do not explicitly consider here).

\begin{table}[bt]
\begin{tabular}{|c||c|c|c|c|c||c|c|c|c|c|}
 \hline
   $m_\chi\to$ &   10 &   30 &  100 &  300 & 1000 &   10 &   30 &  100 &  300 & 1000 \\
\hline
channel & \multicolumn{5}{|c||}{WMAP7 $f_{\rm eff}$} &
\multicolumn{5}{|c|}{Planck $f_{\rm eff}$} \\
 \hline
 $e$        & 0.74 & 0.65 & 0.59 & 0.56 & 0.56 & 0.79 & 0.70 & 0.63 & 0.59 & 0.59 \\
 \hline
 $\mu$      & 0.28 & 0.25 & 0.23 & 0.21 & 0.21 & 0.30 & 0.27 & 0.25 & 0.23 & 0.22 \\
 \hline
 $\tau$     & 0.22 & 0.21 & 0.19 & 0.19 & 0.18 & 0.24 & 0.23 & 0.21 & 0.20 & 0.20 \\
 \hline
 $V\to e$   & 0.77 & 0.69 & 0.62 & 0.56 & 0.56 & 0.83 & 0.74 & 0.66 & 0.60 & 0.60 \\
 \hline
 $V\to\mu$  & 0.28 & 0.26 & 0.23 & 0.21 & 0.20 & 0.30 & 0.28 & 0.25 & 0.23 & 0.21 \\
 \hline
 $V\to\tau$ & 0.22 & 0.22 & 0.20 & 0.18 & 0.18 & 0.24 & 0.23 & 0.21 & 0.20 & 0.19 \\
 \hline
 $q(u,d,s)$        & 0.33 & 0.31 & 0.30 & 0.28 & 0.27 & 0.35 & 0.33 & 0.32 & 0.30 & 0.29 \\
 \hline
 $c$        & 0.33 & 0.32 & 0.30 & 0.28 & 0.27 & 0.36 & 0.34 & 0.32 & 0.30 & 0.29 \\
 \hline
 $b$        & 0.34 & 0.32 & 0.31 & 0.29 & 0.27 & 0.36 & 0.34 & 0.33 & 0.31 & 0.29 \\
 \hline
 $t$        & $-$ & $-$ & $-$ & 0.27 & 0.26 & $-$ & $-$ & $-$ & 0.29 & 0.28 \\
 \hline
 $\gamma$   & 0.58 & 0.50 & 0.59 & 0.56 & 0.55 & 0.62 & 0.54 & 0.62 & 0.60 & 0.58 \\
 \hline
 $g$        & 0.33 & 0.32 & 0.30 & 0.28 & 0.27 & 0.35 & 0.34 & 0.32 & 0.30 & 0.29 \\
 \hline
 $W$        & $-$ & $-$ & 0.26 & 0.25 & 0.24 & $-$ & $-$ & 0.28 & 0.26 & 0.25 \\
 \hline
 $Z$        & $-$ & $-$ & 0.25 & 0.23 & 0.22 & $-$ & $-$ & 0.27 & 0.25 & 0.24 \\
 \hline
 $h$        & $-$ & $-$ & $-$ & 0.28 & 0.26 & $-$ & $-$ & $-$ & 0.30 & 0.28 \\
 \hline
\hline
   $m_\chi\to$ & 40   &   50 &  60 &  70 & 80 &   40 &   50 &  60 & 70 & 80 \\
\hline
$\gamma$ & 0.48 &  0.46 &  0.44 & 0.49 & 0.526 &  0.51 & 0.49 & 0.49 & 0.52 & 0.56 \\
\hline
\end{tabular}
\caption{$f_{\rm eff}$ values as a function of WIMP mass $m_\chi$ (in GeV) and primary
annihilation channel (for example ``$e$'' denotes $\chi\chi\to 
e\bar e$, while ``$V\to e$'' denotes $\chi\chi\to VV$, followed by 
$V\to e\bar e$), for computing WMAP7 (left) and projected Planck 
(right) constraints.
Null entries indicate that $m_\chi$ is below threshold.  Bottom row
covers extra range of masses where $\chi\chi\to\gamma\gamma$ constraints undergo a
transition in fig.\ \ref{fig1}. }
\label{tab1}
\end{table}

\begin{figure}[tb]
\centerline{
\includegraphics[width=0.4\textwidth]{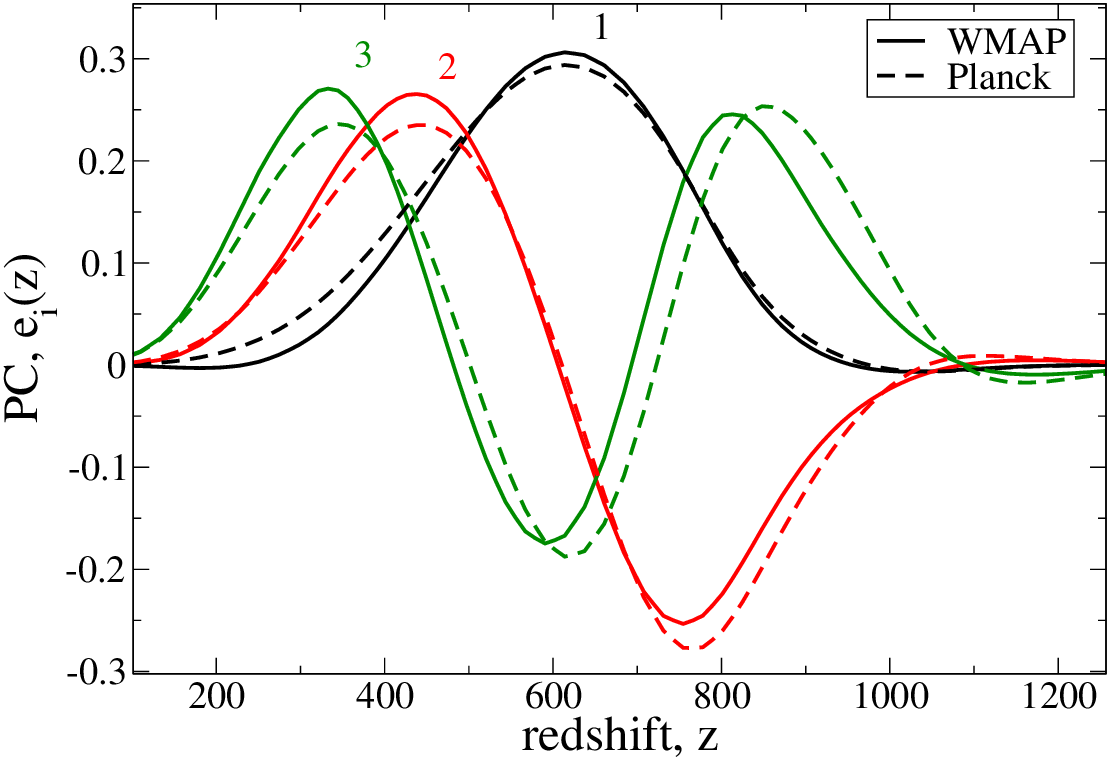}
}
\caption{First three principal components for annihilating dark
matter, from \href{http://nebel.rc.fas.harvard.edu/epsilon/}{http://nebel.rc.fas.harvard.edu/epsilon/}.  Solid curves are for WMAP7 and dashed for anticipated
Planck data.}
\label{fig0}
\end{figure}

\begin{figure*}[bt]
\centerline{
\includegraphics[width=0.5\textwidth]{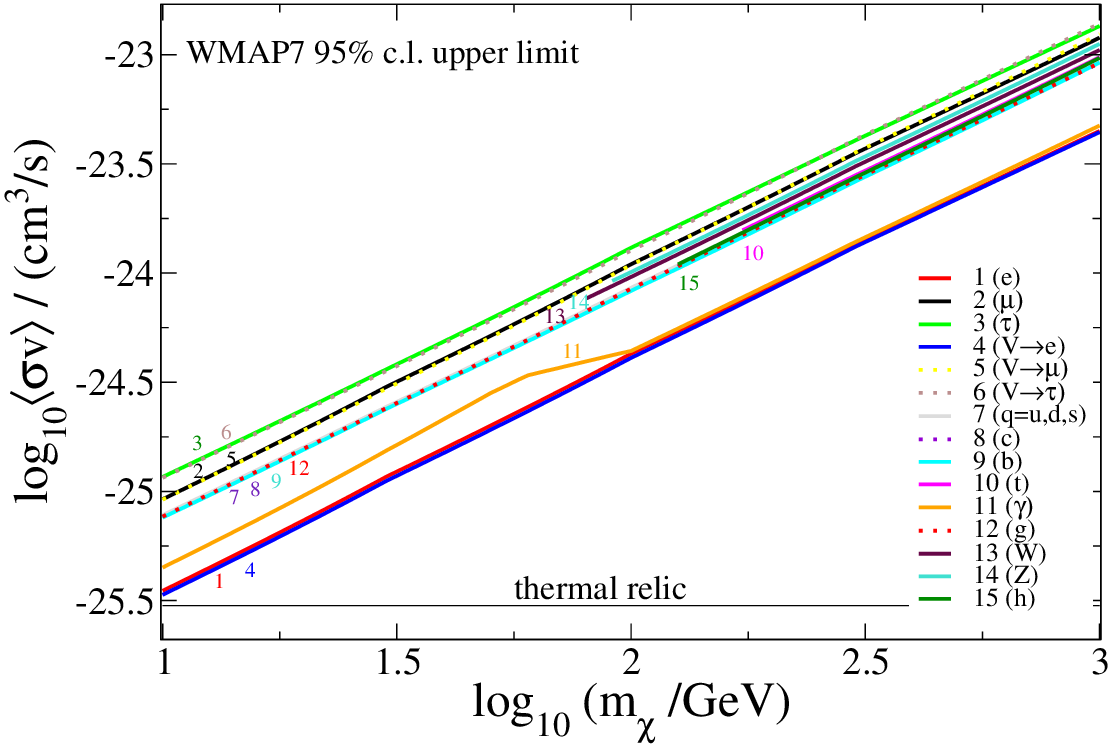}
\includegraphics[width=0.5\textwidth]{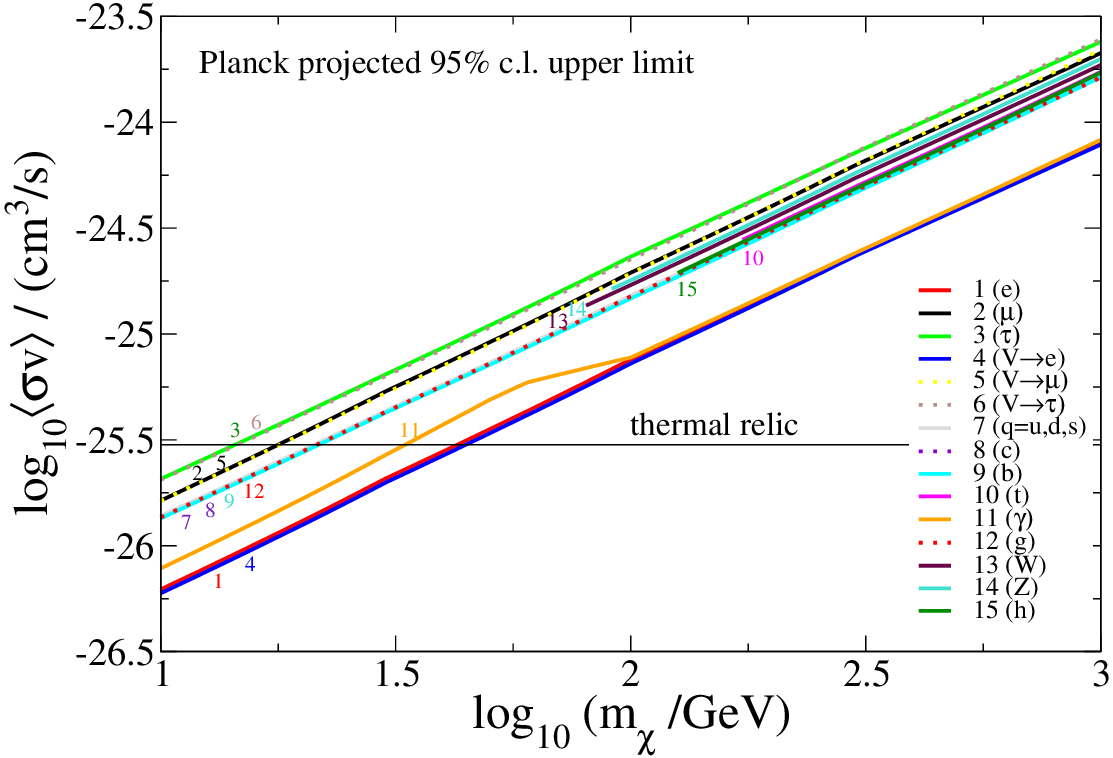}}
\caption{95\% confidence level (CL) limits on cross section 
$\langle\sigma v\rangle$ for annihilation into the 15 channels
indicated, for WMAP7 (left, corresponding to (\ref{wmapsiglim})) and anticipated Planck results (right),
for 14 months of observing time.
As indicated by the numerals, the limiting curves fall into 
nearly-coinciding groups (1,4), 11, (7,8,9,12,10,15), (13,14), (2,5),
(3,6), according to the channel-numbering key indicated on the
figures.  Horizontal line
shows the nominal thermal relic density cross section, $\langle\sigma v\rangle =
3\times 10^{-26}$cm$^3$/s.}
\label{fig1}
\end{figure*}

To find the spectra $\phi_i$, Monte Carlo computations using an event
generator such as PYTHIA \cite{Bengtsson:1987kr, Sjostrand:2007gs} or HERWIG \cite{Corcella:2000bw} are required.  These
simulations are numerically intensive, but the results have been
carried out and made available for a range of primary annihilation and
decay channels in ref.\ \cite{Cirelli:2010xx}.  {The spectra obtained from PYTHIA 8.135 and HERWIG 6.510
generally agree to within about 20\% for most primary annihilation/decay channels, except for $g$ (gluons), where the choice of event generator 
results in up to a factor of two uncertainty in electron and photon yields.  At lower energies PYTHIA also gives substantially larger 
photon yields for leptonic and gauge boson channels, a consequence of 
the omission of (QED) lepton final state radiation and fermion pair production in HERWIG.  We therefore exclusively employ the 
PYTHIA results of ref.\ \cite{Cirelli:2010xx}, which have the additional benefit of also including electroweak corrections \cite{Ciafaloni}.}

The purpose of the present note is to provide a set of results that
can be easily interpolated to find CMB constraints and
likelihood functions for dark matter models that annihilate or decay
to arbitrary final states with a constant cross-section or decay
rate.  We first carry out the computation of (\ref{fzcurve}) for a
full range of dark matter masses and annihilation and decay channels, and
then translate each result into a single number that encodes the energy injection
history for that mass and final state.  For annihilation we use $f_{\rm eff}$, the effective efficiency of energy injection, whereas for decays we define an alternative quantity $\eta$.  We then detail how likelihoods and
corresponding constraints are obtained as a function of $f_{\rm eff}$ or $\eta$,
including a simple prescription for combining results from multiple
channels.  

Here we explicitly build upon existing public datasets
\cite{Finkbeiner:2011dx,Slatyer:2012yq,Cirelli:2010xx}, adding the
necessary final steps for earlier results to be immediately
implemented in, for example, multi-messenger dark matter analyses
\cite{Pato:2009fn} and global fits to beyond the Standard Model
particle theories \cite{Bechtle:2012zk, Buchmueller:2012hv,
Strege:2012bt}.  This complements and extends recent efforts to
construct public likelihood functions from \emph{Fermi}-LAT
\cite{Scott:2009jn}, HESS \cite{Ripken:2010ja} and IceCube
\cite{Scott:2012mq} searches for dark matter annihilation.

{\bf Annihilating dark matter.}  We will first consider the
case of dark matter annihilating into standard model particles.
Bounds on annihilation cross sections can be encoded in
an integral involving $f(z)$ and a set of {\it principal
component} basis functions $e_i(z)$ that should be optimized for
annihilations, and which depend 
somewhat upon
which experiment is being considered since they are eigenvectors of the Fisher matrix.  As discussed in 
\cite{Finkbeiner:2011dx,Slatyer:2012yq}, $f(z)$ can be expanded
as
\be
	\varepsilon\,f(z) = \sum_{i=1}^\infty{\varepsilon_i}e_i(z)
\label{fz2}
\ee
where $\varepsilon\equiv \langle\sigma v\rangle/m_\chi$ and 
$\varepsilon_i = \varepsilon\,f(z)\!\cdot\!\,e_i(z)/
e_i(z)\!\cdot\!\,e_i(z)$.  The inner product is the integral over 
$z$ using the integration limits $z_1=86.83$, $z_2=1258.2$.
  For annihilating dark matter, these basis functions 
are chosen to maximize sensitivity to
a general expected $z$-dependence for energy injection from annihilating dark matter,
(based upon the energy density 
$(1+z)^3\Omega_{\rm DM}\rho_c c^2$ of the dark matter itself)
 in the sense that the most important contributions
to the observable energy deposition are described by the lowest 
components.  The first three $e_i(z)$ are plotted in fig.\ \ref{fig0}.

In fact, the contribution of the first component
has been demonstrated to dominate, especially in the case of WMAP
data, in the computation of the likelihood function
that is relevant for constraining annihilating dark matter models.  
Under the approximation of a Gaussian likelihood, and that the
response of the CMB to the energy injection is linear, 
the chi-squared is given by \cite{Finkbeiner:2011dx,Slatyer:2012yq}
\be
	\Delta\chi^2 = \sum_i\lambda_i \varepsilon_i^2/\bar\varepsilon^2
\label{dc2}
\ee
where $\lambda_i$ is the eigenvalue of the Fisher matrix corresponding to $e_i$,
for the relevant CMB experiment, and $\bar\varepsilon$ is a fiducial value,
taken to be $2\times 10^{-27}$ cm$^3\,$s$^{-1}\,$GeV$^{-1}$.  
For WMAP7, $\lambda_1 = 0.279$, while for Planck, $\lambda_i = 3.16,\, 0.691,\, 0.162\dots$,
and the inclusion of contributions from the first three principal components
is justified.
Then, for example, the $2\sigma$ upper limit on the cross section
is given by
\be
	\langle \sigma v\rangle < {2m_\chi\bar\varepsilon\over
	\sqrt{\sum_i\lambda_i ((f\cdot e_i)/(e_i\cdot e_i))^2}}
\label{siglim}
\ee
The numerical value of the numerator in (\ref{siglim}) (as well as
that of (\ref{wmapsiglim}) below) is
correlated with the choice of normalization of the $e_i(z)$'s.
We use the functions shown in fig.\ \ref{fig0}, which are normalized 
such that $(e_i\cdot e_i)=23.9$.

However, it was observed in ref.\ \cite{Finkbeiner:2011dx} that the approximation of 
linear response is not yet very accurate for the magnitudes of
$\varepsilon_i$ that are allowed by the WMAP7 data; instead one should
perform a full likelihood analysis using \texttt{CosmoMC} \cite{Lewis:2002ah} (and it is found that
only the first principal component is required).  This gives a
somewhat weaker constraint than (\ref{siglim}); the 95\% c.l.\ upper
bound is
\be
	\langle \sigma v\rangle < 1.2\times 10^{-26}{{\rm
	cm}^3\over {\rm s\cdot GeV}}\, {m_\chi\over (f\cdot e_1)/(e_1\cdot e_1)}
\label{wmapsiglim}
\ee
(Throughout this paper, we use ``95\% c.l.'' to mean
$2\sigma$, which is in fact 95.45\% c.l..)

In either (\ref{siglim}) or (\ref{wmapsiglim}), we see that the
constraints can be expressed in terms of a single number involving
integrals of $f(z)$.  To make contact with the earlier literature
\cite{Slatyer:2009yq}, we find it useful to consider a quantity $f_{\rm
eff}$ defined in terms of a ``universal WIMP annihilation'' curve
$\ew(z)$ \cite{Finkbeiner:2011dx}, $f_{\rm eff}\sim (f\cdot\ew)/(\ew\cdot\ew)$, which has the
interpretation that $f_{\rm eff}<1$ denotes the average efficiency
of energy injection for the annihilation channel of interest.
It can be compared to the $f_{\rm mean}$ values tabulated in ref.\ 
\cite{Slatyer:2009yq} (column 2, table I).  Here instead of
defining $f_{\rm eff}$ directly in terms of $\ew$, we use the 
expansion $\ew = \sum_i c_i e_i$, and the observation that using just
the first term in the expansion gives a good approximation.  Thus for WMAP7 we define
\be
	f_{\rm eff} \equiv {(f\cdot e_1)\over c_1
	(e_1\cdot e_1)}\qquad {\rm (WMAP)}
\label{feffwmap}
\ee
where numerically $c_1=4.64$.  Then (\ref{wmapsiglim}) can be expressed as 
\be
\langle\sigma v\rangle < 0.26\times10^{-26}\,{\rm cm}^3\,{\rm s}^{-1}\,{\rm GeV}^{-1} m_\chi/ f_{\rm eff}.
\ee 
The analogous definition
for Planck, which makes use of the contributions from the first three
principal components, is
\be
	f_{\rm eff} \equiv {1\over\sqrt{\lambda_1}c_1}
	\left[\sum_i\lambda_i\left({f\cdot e_i\over e_i\cdot e_i}
	\right)^2\right]^{1/2}\quad ({\rm Planck})
\label{feffplanck}
\ee
where we must use the $\lambda_i$ and $e_i$ appropriate to
Planck.  Then the 2$\sigma$ projected constraint from Planck 
takes the form  $\langle\sigma v\rangle < 0.48\times10^{-27}\,{\rm
cm}^3\,{\rm s}^{-1}\,{\rm GeV}^{-1} m_\chi/ f_{\rm eff}$.  
For annihilation into several channels $i$ with branching
fractions $r_i$, the total $f_{\rm eff}$ is
\be
 f_{\rm eff} = \sum_i r_i f_{{\rm eff},i}.
\ee

We present the 95\% CL constraints as a function of $m_\chi$ for
a range of annihilation 
channels in fig.\ \ref{fig1}, interpolating from table \ref{tab1}. 
The final states are pairs of $e$, $\mu$, $\tau$, $V\to e,$\footnote{$V\to
X$ denotes
$\chi\chi\to VV$ followed by $V\to X\bar X$ with $m_V=1$\,GeV for
$X=e$ or $\mu$ and $m_V=5$\,GeV for $X=\tau$.}
$V\to\mu$, $V\to\tau$, $q$ (light quarks: $u,d$ or $s$), 
$c$, $b$, $t$, $W$, $Z$, $g$ (gluons), $\gamma$, or $h$ (Higgs).
The limits from
WMAP7 and projected constraints from Planck based on 
14 months of observation time \cite{Planck:2006aa} are shown in the figure.  Our 95\% C.L. WMAP7 limits perfectly match those for $e$ and $\mu$ found in present benchmark results (e.g.\ at $m_\chi=100$\,GeV, $\langle\sigma v\rangle < 4\times10^{-25}$\,cm$^{3}$\,s$^{-1}$ for $e^+e^-$, $\langle\sigma v\rangle <1\times10^{-24}$\,cm$^{3}$\,s$^{-1}$ for $\mu^+\mu^-$ \cite{Galli:2011rz, Finkbeiner:2011dx}).  We see approximately a factor of 6 improvement between WMAP7 and \textit{Planck}, in agreement with the results of \cite{Hutsi:2011vx} but slightly less than the factor of 8 reported by \cite{Galli:2011rz}.

Limits are strongest on the channels $[e,\,V\to e]$ (which in fact coincide) and $\gamma$, which 
also coincides with the former at masses $\gtrsim 100$\,GeV.  The $\gamma\gamma$ final state limit has a different shape from the
others because of the monochromatic nature of the photons and the complex dependence of the efficiency of photon
energy deposition on redshift and injection energy.  This is encoded in the transfer function $\tilde T_\gamma$, and can be seen for example in fig.\ 1 of \cite{Slatyer:2012yq} and fig.\ 2 of \cite{Chen:2003gz}.  At redshifts of $z\sim600$, where the first principal component of fig.\ \ref{fig0} is peaked, the Universe is approximately transparent to photons with energies $E\lesssim50\,$GeV, but rapidly transitions to completely opaque for $E\gtrsim100\,$GeV, due to scattering and pair production on CMB photons \cite{Chen:2003gz}.  This causes the $\gamma\gamma$ curve to track the electron curve above 100\,GeV, as in both cases essentially all of the annihilation energy goes into heating the primordial gas.  The $\gamma\gamma$ channel is the only one in which photons dominate over electrons in the final state spectra, and the only one to produce monochromatic photons -- so it is the only one to show this complicated behaviour in the limits.

The next strongest constraints are on the channels involving 
quarks, gluons and Higgs.  These all coincide with each other
(except that the top quark and Higgs have a higher mass threshold).
The next weakest constraints are on $[W,\, Z]$, followed by 
$[\mu,\, V\to\mu]$ and finally $[\tau,\, V\to\tau]$.  These tendencies
are indicative of the greater fraction of energy ending up in
neutrinos for the least constrained channels.

We stress that our results do not depend in any intrinsic way upon the
somewhat arbitrary definitions (\ref{feffwmap},\ref{feffplanck}).
Any quantity proportional to $f\cdot e_1$ in the case of WMAP or
$(\sum_i\lambda_i(f\cdot e_i/e_i\cdot e_i)^2)^{1/2}$ in 
the case of Planck would
suffice to encode the necessary information for recovering the 
constraints.  To this end, we list $f_{\rm eff}$ values for both WMAP
and Planck in table \ref{tab1},  for the annihilation channels
mentioned above and for the WIMP masses 
$m_\chi = 10,\, 30,\, 100,\, 300,\, 1000$\,GeV. They are in 
reasonable agreement with the $f_{\rm mean}$ values tabulated 
in ref.\ \cite{Slatyer:2009yq} in the cases that overlap.

To determine constraints at an arbitrary confidence level, one would
like to know the likelihood function for the annihilation cross
section, assuming branching fractions $r_i$ to channel $i$.  For Planck, this is
given by
\be
\begin{split}
 \ln\, &\mathcal{L}(\langle\sigma v\rangle|m_\chi,r_i) = \\
 &-\frac12 f_{\rm eff}^2(m_\chi,r_i) \lambda_1 c_1^2\,
 \left(\frac{\langle\sigma v\rangle}{2\times 10^{-27}{\rm cm}^3{\rm s}^{-1}}\right)^2\,\left(\frac{\rm GeV}{m_\chi}\right)^2.\\
\label{loglike2}
\end{split}
\ee             
from (\ref{siglim}) and (\ref{feffplanck}).  As mentioned above, this
expression, which assumes linear response of the CMB to the deposited
energy, is not accurate for WMAP.   This can be corrected by making the
replacement {$2\times 10^{-27}\to 3.2\times 10^{-27}$} in 
(\ref{loglike2})  (and one must also use the appropriate value
$\lambda_1=0.279$ for WMAP).  Similarly, caution should be exercised if (\ref{loglike2}) is used to compute WMAP likelihoods far from
95\% c.l.\, as the Gaussian approximation was observed
to be poor for WMAP likelihoods in ref.\ \cite{Finkbeiner:2011dx}.  
{The WMAP 7-year limits can also be improved by $\sim$15\% if ACT data are added
to the \texttt{CosmoMC} fit \cite{Galli:2011rz}, and presumably slightly more than this if SPT data are added.}

\begin{figure}[tb]
\centerline{
\includegraphics[width=0.4\textwidth]{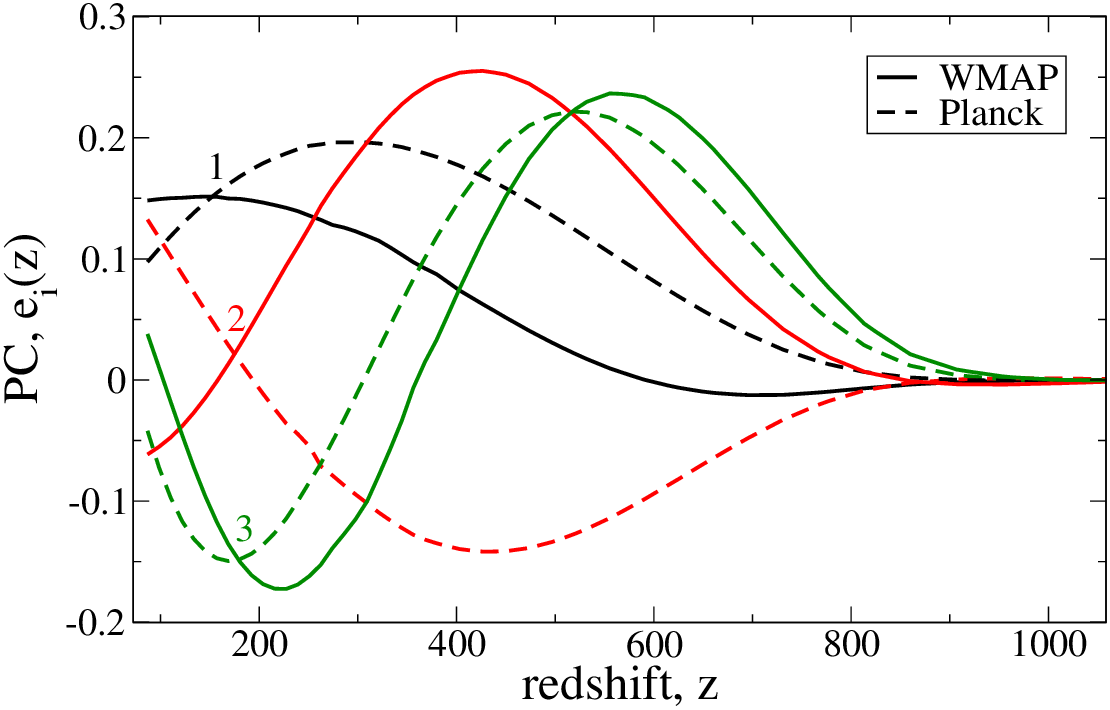}
}
\caption{First three principal components for decaying dark
matter, from \href{http://nebel.rc.fas.harvard.edu/epsilon/}{http://nebel.rc.fas.harvard.edu/epsilon/}.  Solid curves are for WMAP7 and dashed for anticipated
Planck data.}
\label{fig2}
\end{figure}

\begin{figure}[tb]
\centerline{
\includegraphics[width=0.45\textwidth]{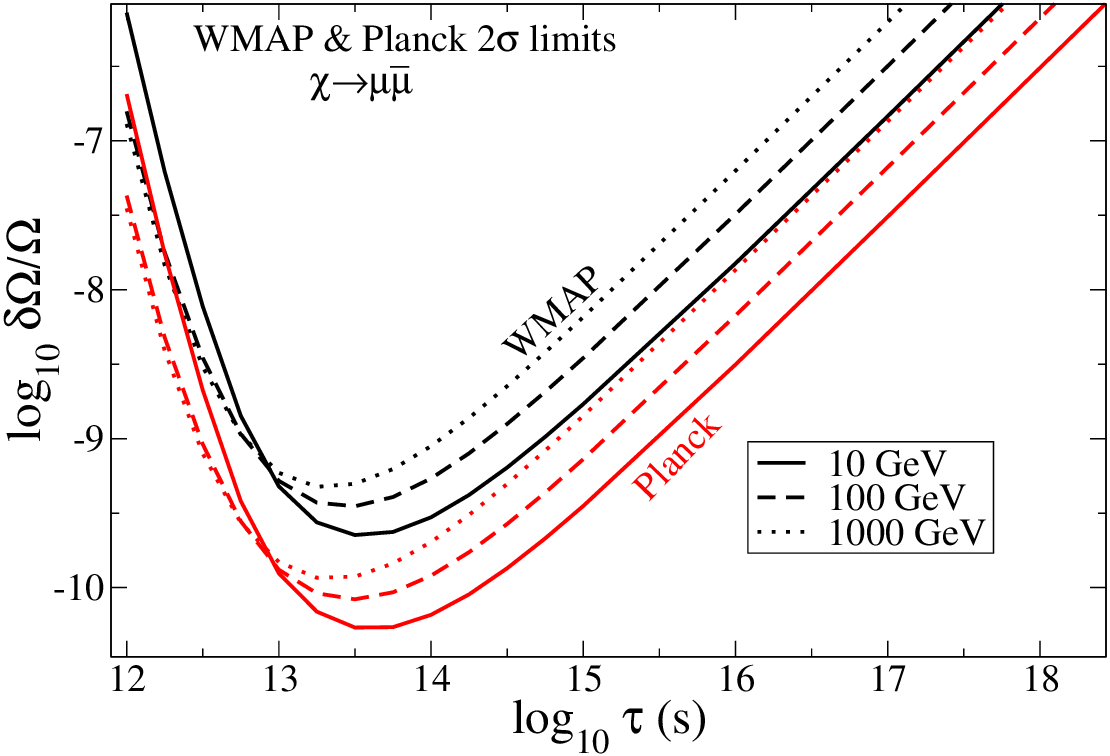}
}
\caption{95\% c.l.\ constraints on $\delta\Omega/\Omega_\DM$ versus $\tau$
for $\chi\to \mu\bar \mu$ decays, with 
$m_\chi = 10, 100, 1000$\,GeV (solid, dashed, dotted lines
respectively), from WMAP7 (upper curves) or Planck (lower curves).}
\label{fig3}
\end{figure}

\begin{figure*}[tb]
\centerline{
\includegraphics[width=0.45\textwidth]{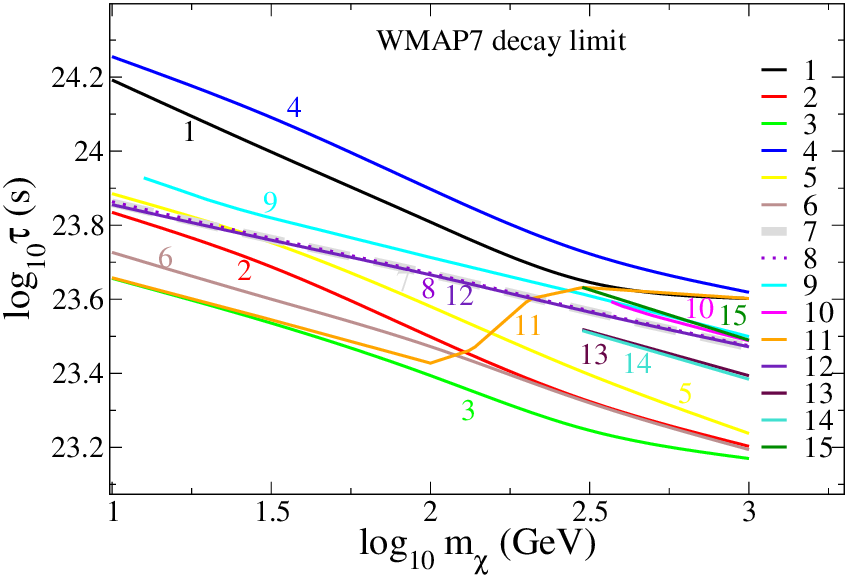}$\ $
\includegraphics[width=0.45\textwidth]{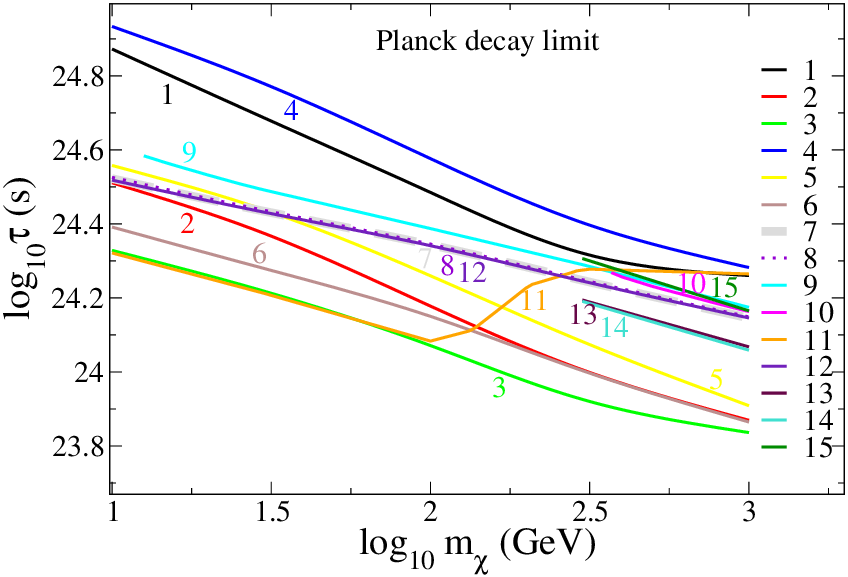}
}
\caption{95\% c.l.\ lower bounds on the lifetime of decaying DM
that accounts for the total DM density, 
from WMAP7 (left) or projected Planck data (right).  The numbering
of the final state channels is the same as in fig.\ \ref{fig2}:
1:$e$, 2:$\mu$, 3:$\tau$, 4:$(V\to e)$,
5:$(V\to\mu)$, 6:$(V\to\tau)$, 7:$(q=u,d,s)$, 8:$c$, 9:$b$, 10:$t$, 
11:$\gamma$, 12:$g$, 13:$W$, 14:$Z$, 15:$h$.}
\label{fig4}
\end{figure*}

{\bf Decaying dark matter.}  A similar procedure can be used to
constrain the fractional mass abundance $\delta\Omega$ of some metastable species present at the time of recombination, which
decays into the same set of final states as we assumed above for
annihilations.  If the decaying species originally contributes
$\Omega_i$ to the mass density of the universe relative to $\rho_c$, and if the decay of a particle of that species converts a fraction
$\phi$ of its rest mass to standard model particles 
(and a fraction $1-\phi$ to some lighter dark species), then $\delta\Omega = \phi\Omega_i$. The injected electromagnetic power density goes as 
\be	
{dE\over dt\,dV} = 
{f(z)\over\tau}\, e^{-t/\tau}\,\delta\Omega\, \rho_c\,c^2\,(1+z)^3.
\ee
The factor of $(1+z)^3\sim n_\chi$ as opposed to 
$(1+z)^6\sim n_\chi^2$ for annihilations is the reason for the
different powers of $(1+z)$ appearing in footnote \ref{fn1}.
Unlike the case of annihilation, where the cross section appears only
as a prefactor, here we have dependence upon the lifetime $\tau$ not
only in the prefactor, but also in the $z$-dependent function
$e^{-t(z)/\tau}$, which is not present in (\ref{power}).  This makes
the analysis of decays more cumbersome than that of annihilations, because the
transfer functions $\tilde T_i(z',z,E)$ now also depend upon $\tau$.  We therefore 
need to compute (\ref{fzcurve}) not only for all masses and final states, but for a 
grid of lifetimes as well.

Note that this form of analysis allows us to not only constrain the
lifetime and abundance of decaying dark matter candidates, but also
the initial abundance of other metastable species, with such short
lifetimes that they would not contribute to the current-day abundance
of dark matter.  For lifetimes less than $\sim$$10^{17}$--$10^{18}$\,s
the particles we refer to are therefore not dark matter in the usual
sense, but rather
some generic metastable particles.

In the approximation of linear response to the injected energy,
the chi-squared is most conveniently written in the form 
\cite{Slatyer:2012yq} 
\be
	\Delta\chi^2 = \left(\delta\Omega\over\Omega_\DM\tau\right)^2\eta^2,
\label{decaychi2}
\ee
where we define 
\be
\label{eta}
\eta=\left[\sum_i \left({(f\cdot e_i)\over (e_i\cdot e_i)\,\sigma_i}\right)^2\right]^{1/2}
\ee 
to be the single number that encapsulates the energy injection history for a given mass, lifetime and final state (as $f_{\rm eff}$ does for annihilation).  The likelihood is then simply
\be
\label{declike}
\ln\, \mathcal{L}(\tau|m_\chi,r_i) = -\frac12\left(\delta\Omega\over\Omega_\DM\tau\right)^2\eta^2(\tau,m_\chi,r_i).
\ee

The principal components $e_i(z)$ in $\eta$ are now specialized to the
case of decays, as well as depending upon the experiment.  In contrast to
annihilations, the first three components are needed for good
accuracy, even for WMAP.  We plot them for both experiments in fig.\ \ref{fig2}. 
The variances are given by $\sigma_i = 
(5.3,\,6.4,\,9.5\dots)\times 10^{-25}\,$s$^{-1}$ for WMAP7 and
$(1.2,\,1.6,\,2.3\dots)\times 10^{-25}\,$s$^{-1}$ for Planck \cite{Slatyer:2012yq}. The limits
of integration appearing in $f\cdot e_i$ are $z_1=10$, $z_2=1258.2$, 
because $n_\chi$ redshifts slower than $n_\chi^2$, so lower redshifts
are needed than in the case of annihilations.  The inner products 
(and therefore the limits of integration) are now defined in terms of integrals over
$\ln z$ rather than $z$, with normalization $(e_i\cdot e_i) = 0.0543$.

Like $f_{\rm eff}$, $\eta$ can be easily obtained for arbitrary branching fractions $r_i$ into multiple final states by taking the appropriate linear combination of values for single \mbox{channels $\eta_i$}
\be
 \eta = \sum_i r_i \eta_i.
\ee

We have computed $\eta$ for each of the 15 final
states that were considered as annihilation channels in the 
previous section, over a grid of metastable particle masses $10, 30, 100, 300, 1000$\,GeV
as before (we augment this selection in the $\gamma$, $t$, $W$, $Z$
and $h$ channels where it would otherwise be too sparse),
 and a grid of 15 lifetimes.  The results are given in table \ref{tab2}.  
{For the $b$ channel, our results extend only as low as $m_\chi=12$\,GeV, as all 
yields of \cite{Cirelli:2010xx} for the $b$ channel go to zero as the energies of the the decay
products approach 5\,GeV (presumably because the $b$ mass was approximated to 5\,GeV in the calculations of \cite{Cirelli:2010xx}).}

Our lifetime grid is sufficiently dense to
show the structure of the limit on $\delta\Omega$ as a function of 
$\tau$.  For longer lifetimes, $\eta$ is the same as the values we give for $\tau=10^{17}$, allowing constraints and likelihoods to be extended to arbitrarily long lifetimes.  An example of the resulting constraints is given in fig.\
\ref{fig3} for the $\chi\to \mu\bar \mu$ channel.  Our results for other channels agree with those presented by Slatyer \cite{Slatyer:2012yq} (as expected, given that we adopt her transfer functions here).  In fig.\ \ref{fig4}
we plot the lower limit on the lifetime of a dark matter
species decaying into any of the channels under consideration, 
assuming that it constitutes the entire presently-observed DM density.

We again added extra points to the mass grid for the $\gamma$ channel, in order
to resolve the somewhat sharp feature seen in the region
$100\lesssim m_\chi\lesssim300$\,GeV, as discussed previously in the results for
annihilating dark matter.  Note here that the mass range of the transition is approximately a factor of two larger than seen in annihilation (as decays produce photons with energies equal to half the DM mass), but that this factor is not exact because the redshifts of importance differ in the two cases, so the energy above which the Universe is opaque to photons injected at relevant redshifts differs slightly.

Our results for decaying species are based on the Fisher-matrix
analysis of \cite{Finkbeiner:2011dx, Slatyer:2012yq}, which assumes
that the linear approximation for the response of the CMB to energy
injection at high redshift holds.  Given the small ionization
fraction at the redshifts most important for decay (at least when the lifetime is long) we do not expect nonlinear
corrections to be crucial.  In principle though, this
assumption needs to be verified, for example with explicit \texttt{CosmoMC}
calculations, as was done for annihilation.

It is also worth noting that the principal components $e_i(z)$ that we use for decay have been
optimized for long decay lifetimes, leading to large leading contributions at relatively 
low redshift (fig.\ \ref{fig2}).  In principle the $e_i(z)$ could also be optimized
for different decay lifetimes, such that the optimal components for shorter lifetimes
would probably begin to resemble those for annihilation to some degree (fig.\ \ref{fig0}),
potentially allowing the limits we give here to be improved for the shortest lifetimes.

With tables \ref{tab1} and \ref{tab2}, and eqs.\ \ref{loglike2} and
\ref{declike}, we have provided a fast and easy means for 
implementing CMB
constraints and likelihoods in future analyses of
annihilating and decaying dark matter models, as well as some models with metastable particles that do not constitute dark matter.   Our results can be
interpolated to provide likelihood functions and limits for
arbitrary particle masses, and arbitrary annihilation and decay
final state mixtures.

\begin{table*}[bt]
\centering \tiny
\begin{tabular}{|c|r||c|c|c|c|c|c|c|c|c|c|c|c|c|c|c||c|c|c|c|c|c|c|c|c|c|c|c|c|c|c|}
\hline
\multicolumn{2}{|r||}{$\log\tau$} &12& 12$\frac14$& 12$\frac12$ &12$\frac34$& 13& 13$\frac14$& 13$\frac12$& 13$\frac34$& 14 &
14$\frac14$& 14$\frac12$& 14$\frac34$& 15&16&17& 12& 12$\frac14$& 12$\frac12$ &12$\frac34$& 13& 13$\frac14$& 13$\frac12$& 13$\frac34$& 14 &
14$\frac14$& 14$\frac12$& 14$\frac34$& 15&16&17\\
 \hline
ch& $m_\chi$ & \multicolumn{15}{|c||}{WMAP7 $\eta$} &
\multicolumn{15}{|c|}{Planck $\eta$} \\
\hline
1&    10 & 5.74 & 4.51 & 3.47 & 2.59 & 1.93 & 1.47 & 1.15 & 0.93 & 0.79 & 0.69 & 0.63 & 0.59 & 0.56 & 0.51 & 0.51 & 5.19 & 3.95 & 2.90 & 2.01 & 1.34 & 0.86 & 0.53 & 0.29 & 0.13 & 0.02 & -.05 & -.09 & -.12 & -.17 & -.17\\
&    30 & 5.49 & 4.29 & 3.32 & 2.54 & 1.96 & 1.55 & 1.27 & 1.07 & 0.94 & 0.85 & 0.80 & 0.76 & 0.74 & 0.70 & 0.69 & 4.92 & 3.72 & 2.75 & 1.96 & 1.36 & 0.94 & 0.64 & 0.43 & 0.28 & 0.19 & 0.12 & 0.08 & 0.06 & 0.02 & 0.01\\
&   100 & 5.39 & 4.20 & 3.26 & 2.53 & 2.00 & 1.63 & 1.37 & 1.20 & 1.09 & 1.02 & 0.97 & 0.94 & 0.92 & 0.89 & 0.89 & 4.82 & 3.63 & 2.68 & 1.94 & 1.40 & 1.02 & 0.75 & 0.56 & 0.44 & 0.36 & 0.30 & 0.27 & 0.25 & 0.22 & 0.21\\
&   300 & 5.38 & 4.18 & 3.24 & 2.54 & 2.04 & 1.69 & 1.46 & 1.31 & 1.21 & 1.15 & 1.11 & 1.09 & 1.07 & 1.05 & 1.05 & 4.81 & 3.60 & 2.66 & 1.95 & 1.44 & 1.08 & 0.84 & 0.67 & 0.56 & 0.49 & 0.45 & 0.42 & 0.40 & 0.38 & 0.38\\
&  $10^3$ & 5.38 & 4.18 & 3.25 & 2.54 & 2.04 & 1.70 & 1.47 & 1.33 & 1.23 & 1.18 & 1.14 & 1.12 & 1.11 & 1.10 & 1.10 & 4.81 & 3.61 & 2.67 & 1.95 & 1.44 & 1.09 & 0.85 & 0.69 & 0.59 & 0.53 & 0.49 & 0.47 & 0.46 & 0.44 & 0.44\\
 \hline
2&    10 & 6.56 & 5.24 & 4.09 & 3.10 & 2.38 & 1.89 & 1.55 & 1.32 & 1.17 & 1.07 & 1.01 & 0.96 & 0.93 & 0.88 & 0.86 & 6.01 & 4.69 & 3.52 & 2.53 & 1.79 & 1.29 & 0.93 & 0.68 & 0.52 & 0.40 & 0.33 & 0.28 & 0.25 & 0.20 & 0.19\\
&    30 & 6.06 & 4.85 & 3.85 & 3.01 & 2.38 & 1.94 & 1.63 & 1.42 & 1.28 & 1.18 & 1.12 & 1.08 & 1.06 & 1.01 & 1.01 & 5.50 & 4.29 & 3.28 & 2.43 & 1.79 & 1.33 & 1.01 & 0.78 & 0.62 & 0.52 & 0.45 & 0.40 & 0.37 & 0.33 & 0.33\\
&   100 & 5.90 & 4.70 & 3.74 & 2.98 & 2.42 & 2.02 & 1.75 & 1.56 & 1.43 & 1.35 & 1.30 & 1.26 & 1.24 & 1.21 & 1.20 & 5.33 & 4.13 & 3.16 & 2.39 & 1.82 & 1.41 & 1.12 & 0.92 & 0.78 & 0.69 & 0.63 & 0.59 & 0.56 & 0.53 & 0.52\\
&   300 & 5.82 & 4.63 & 3.69 & 2.97 & 2.45 & 2.08 & 1.84 & 1.67 & 1.56 & 1.49 & 1.44 & 1.42 & 1.40 & 1.37 & 1.37 & 5.26 & 4.05 & 3.11 & 2.38 & 1.85 & 1.47 & 1.21 & 1.03 & 0.91 & 0.83 & 0.78 & 0.75 & 0.72 & 0.70 & 0.69\\
&  $10^3$ & 5.81 & 4.61 & 3.68 & 2.98 & 2.47 & 2.13 & 1.90 & 1.74 & 1.65 & 1.59 & 1.55 & 1.53 & 1.52 & 1.50 & 1.50 & 5.24 & 4.04 & 3.10 & 2.38 & 1.87 & 1.51 & 1.27 & 1.11 & 1.01 & 0.94 & 0.89 & 0.87 & 0.85 & 0.83 & 0.83\\
 \hline
3&    10 & 6.18 & 4.96 & 3.95 & 3.11 & 2.47 & 2.03 & 1.72 & 1.50 & 1.36 & 1.26 & 1.20 & 1.16 & 1.13 & 1.06 & 1.04 & 5.61 & 4.39 & 3.37 & 2.53 & 1.88 & 1.42 & 1.10 & 0.86 & 0.71 & 0.60 & 0.53 & 0.48 & 0.44 & 0.39 & 0.37\\
&    30 & 6.07 & 4.85 & 3.87 & 3.07 & 2.47 & 2.05 & 1.76 & 1.55 & 1.42 & 1.33 & 1.27 & 1.24 & 1.21 & 1.17 & 1.16 & 5.51 & 4.29 & 3.29 & 2.49 & 1.88 & 1.44 & 1.13 & 0.92 & 0.77 & 0.67 & 0.60 & 0.56 & 0.53 & 0.49 & 0.48\\
&   100 & 5.93 & 4.73 & 3.78 & 3.04 & 2.49 & 2.11 & 1.84 & 1.65 & 1.53 & 1.45 & 1.40 & 1.37 & 1.34 & 1.31 & 1.31 & 5.36 & 4.16 & 3.20 & 2.45 & 1.90 & 1.50 & 1.21 & 1.01 & 0.88 & 0.79 & 0.73 & 0.69 & 0.67 & 0.63 & 0.63\\
&   300 & 5.90 & 4.70 & 3.76 & 3.04 & 2.52 & 2.16 & 1.91 & 1.75 & 1.64 & 1.57 & 1.52 & 1.49 & 1.48 & 1.45 & 1.45 & 5.33 & 4.12 & 3.18 & 2.45 & 1.92 & 1.55 & 1.29 & 1.11 & 0.99 & 0.91 & 0.86 & 0.83 & 0.80 & 0.78 & 0.77\\
&  $10^3$ & 5.88 & 4.68 & 3.75 & 3.03 & 2.52 & 2.17 & 1.94 & 1.78 & 1.69 & 1.62 & 1.59 & 1.56 & 1.55 & 1.53 & 1.53 & 5.31 & 4.11 & 3.16 & 2.44 & 1.92 & 1.56 & 1.32 & 1.15 & 1.04 & 0.97 & 0.93 & 0.90 & 0.89 & 0.87 & 0.86\\
 \hline
4&    10 & 5.98 & 4.70 & 3.58 & 2.63 & 1.93 & 1.45 & 1.12 & 0.89 & 0.74 & 0.64 & 0.58 & 0.53 & 0.50 & 0.45 & 0.44 & 5.42 & 4.14 & 3.02 & 2.06 & 1.34 & 0.84 & 0.49 & 0.25 & 0.09 & -.03 & -.10 & -.15 & -.18 & -.23 & -.23\\
&    30 & 5.56 & 4.36 & 3.37 & 2.56 & 1.94 & 1.51 & 1.21 & 1.00 & 0.87 & 0.77 & 0.72 & 0.68 & 0.65 & 0.61 & 0.60 & 5.00 & 3.79 & 2.80 & 1.98 & 1.35 & 0.90 & 0.59 & 0.36 & 0.21 & 0.11 & 0.04 & -.00 & -.03 & -.07 & -.08\\
&   100 & 5.44 & 4.24 & 3.29 & 2.54 & 1.98 & 1.59 & 1.33 & 1.14 & 1.02 & 0.94 & 0.89 & 0.86 & 0.84 & 0.81 & 0.80 & 4.87 & 3.67 & 2.71 & 1.95 & 1.39 & 0.99 & 0.70 & 0.50 & 0.37 & 0.28 & 0.22 & 0.18 & 0.16 & 0.13 & 0.12\\
&   300 & 5.38 & 4.18 & 3.25 & 2.54 & 2.02 & 1.66 & 1.42 & 1.26 & 1.15 & 1.08 & 1.04 & 1.01 & 1.00 & 0.97 & 0.97 & 4.81 & 3.61 & 2.67 & 1.95 & 1.42 & 1.05 & 0.80 & 0.62 & 0.50 & 0.43 & 0.38 & 0.34 & 0.33 & 0.30 & 0.30\\
&  $10^3$ & 5.38 & 4.18 & 3.25 & 2.54 & 2.04 & 1.70 & 1.47 & 1.32 & 1.23 & 1.17 & 1.13 & 1.11 & 1.10 & 1.08 & 1.08 & 4.81 & 3.61 & 2.66 & 1.95 & 1.44 & 1.08 & 0.85 & 0.69 & 0.58 & 0.52 & 0.48 & 0.45 & 0.44 & 0.42 & 0.42\\
 \hline
5&    10 & 6.82 & 5.41 & 4.18 & 3.14 & 2.39 & 1.89 & 1.54 & 1.31 & 1.15 & 1.05 & 0.98 & 0.93 & 0.90 & 0.83 & 0.81 & 6.28 & 4.87 & 3.62 & 2.57 & 1.81 & 1.29 & 0.92 & 0.67 & 0.50 & 0.38 & 0.30 & 0.25 & 0.21 & 0.16 & 0.14\\
&    30 & 6.22 & 4.99 & 3.95 & 3.06 & 2.38 & 1.92 & 1.60 & 1.38 & 1.23 & 1.13 & 1.07 & 1.03 & 1.00 & 0.95 & 0.94 & 5.66 & 4.43 & 3.37 & 2.48 & 1.80 & 1.32 & 0.98 & 0.74 & 0.58 & 0.47 & 0.39 & 0.35 & 0.32 & 0.27 & 0.26\\
&   100 & 5.97 & 4.77 & 3.79 & 3.01 & 2.41 & 1.99 & 1.70 & 1.51 & 1.37 & 1.29 & 1.23 & 1.19 & 1.17 & 1.13 & 1.12 & 5.40 & 4.20 & 3.22 & 2.42 & 1.82 & 1.39 & 1.08 & 0.87 & 0.72 & 0.62 & 0.55 & 0.51 & 0.48 & 0.44 & 0.44\\
&   300 & 5.87 & 4.67 & 3.73 & 2.99 & 2.44 & 2.07 & 1.80 & 1.63 & 1.51 & 1.43 & 1.38 & 1.35 & 1.33 & 1.30 & 1.30 & 5.30 & 4.10 & 3.15 & 2.40 & 1.85 & 1.46 & 1.18 & 0.99 & 0.86 & 0.77 & 0.71 & 0.68 & 0.65 & 0.62 & 0.62\\
&  $10^3$ & 5.84 & 4.64 & 3.70 & 2.99 & 2.48 & 2.13 & 1.89 & 1.73 & 1.63 & 1.57 & 1.53 & 1.50 & 1.49 & 1.46 & 1.46 & 5.27 & 4.06 & 3.12 & 2.40 & 1.88 & 1.52 & 1.27 & 1.10 & 0.98 & 0.91 & 0.86 & 0.84 & 0.82 & 0.79 & 0.79\\
 \hline
6&    10 & 6.17 & 4.96 & 3.96 & 3.12 & 2.47 & 2.02 & 1.70 & 1.48 & 1.34 & 1.24 & 1.17 & 1.12 & 1.08 & 1.00 & 0.97 & 5.61 & 4.39 & 3.38 & 2.53 & 1.88 & 1.42 & 1.08 & 0.85 & 0.68 & 0.57 & 0.49 & 0.44 & 0.40 & 0.33 & 0.31\\
&    30 & 6.15 & 4.93 & 3.92 & 3.10 & 2.47 & 2.03 & 1.73 & 1.52 & 1.38 & 1.29 & 1.23 & 1.19 & 1.16 & 1.11 & 1.09 & 5.59 & 4.36 & 3.35 & 2.52 & 1.88 & 1.43 & 1.11 & 0.88 & 0.73 & 0.63 & 0.56 & 0.51 & 0.48 & 0.43 & 0.42\\
&   100 & 5.99 & 4.79 & 3.83 & 3.06 & 2.48 & 2.08 & 1.79 & 1.60 & 1.47 & 1.39 & 1.33 & 1.30 & 1.27 & 1.23 & 1.23 & 5.43 & 4.22 & 3.25 & 2.47 & 1.89 & 1.47 & 1.17 & 0.96 & 0.82 & 0.72 & 0.66 & 0.62 & 0.59 & 0.55 & 0.55\\
&   300 & 5.92 & 4.72 & 3.78 & 3.05 & 2.51 & 2.14 & 1.88 & 1.70 & 1.58 & 1.51 & 1.46 & 1.43 & 1.41 & 1.38 & 1.37 & 5.35 & 4.14 & 3.20 & 2.46 & 1.91 & 1.53 & 1.25 & 1.06 & 0.93 & 0.85 & 0.79 & 0.75 & 0.73 & 0.70 & 0.70\\
&  $10^3$ & 5.90 & 4.70 & 3.76 & 3.05 & 2.53 & 2.18 & 1.94 & 1.78 & 1.68 & 1.61 & 1.57 & 1.54 & 1.53 & 1.51 & 1.51 & 5.33 & 4.13 & 3.18 & 2.46 & 1.93 & 1.57 & 1.32 & 1.15 & 1.03 & 0.96 & 0.91 & 0.88 & 0.86 & 0.84 & 0.83\\
 \hline
7&    10 & 5.87 & 4.67 & 3.69 & 2.90 & 2.29 & 1.86 & 1.56 & 1.35 & 1.21 & 1.11 & 1.05 & 1.00 & 0.96 & 0.87 & 0.84 & 5.30 & 4.10 & 3.11 & 2.31 & 1.69 & 1.25 & 0.94 & 0.71 & 0.55 & 0.44 & 0.37 & 0.31 & 0.28 & 0.20 & 0.18\\
&    30 & 5.89 & 4.68 & 3.70 & 2.91 & 2.30 & 1.88 & 1.59 & 1.38 & 1.25 & 1.16 & 1.09 & 1.05 & 1.02 & 0.95 & 0.93 & 5.32 & 4.11 & 3.12 & 2.32 & 1.71 & 1.28 & 0.97 & 0.75 & 0.59 & 0.49 & 0.42 & 0.37 & 0.34 & 0.28 & 0.26\\
&   100 & 5.84 & 4.64 & 3.67 & 2.90 & 2.32 & 1.91 & 1.63 & 1.43 & 1.30 & 1.21 & 1.16 & 1.12 & 1.09 & 1.04 & 1.03 & 5.28 & 4.06 & 3.09 & 2.31 & 1.73 & 1.31 & 1.01 & 0.79 & 0.65 & 0.55 & 0.49 & 0.44 & 0.41 & 0.37 & 0.36\\
&   300 & 5.79 & 4.59 & 3.64 & 2.90 & 2.34 & 1.95 & 1.68 & 1.49 & 1.37 & 1.29 & 1.23 & 1.20 & 1.18 & 1.14 & 1.13 & 5.23 & 4.02 & 3.06 & 2.31 & 1.74 & 1.34 & 1.06 & 0.85 & 0.72 & 0.63 & 0.56 & 0.52 & 0.50 & 0.46 & 0.45\\
&  $10^3$ & 5.77 & 4.56 & 3.62 & 2.89 & 2.36 & 1.98 & 1.73 & 1.55 & 1.44 & 1.36 & 1.32 & 1.29 & 1.27 & 1.23 & 1.23 & 5.20 & 3.99 & 3.04 & 2.30 & 1.76 & 1.38 & 1.11 & 0.92 & 0.79 & 0.70 & 0.65 & 0.61 & 0.59 & 0.56 & 0.55\\
 \hline
8&    10 & 5.85 & 4.65 & 3.68 & 2.88 & 2.28 & 1.85 & 1.55 & 1.34 & 1.20 & 1.11 & 1.04 & 0.99 & 0.96 & 0.86 & 0.84 & 5.29 & 4.08 & 3.10 & 2.30 & 1.68 & 1.25 & 0.93 & 0.71 & 0.55 & 0.44 & 0.36 & 0.31 & 0.27 & 0.20 & 0.18\\
&    30 & 5.88 & 4.67 & 3.69 & 2.90 & 2.30 & 1.87 & 1.58 & 1.38 & 1.24 & 1.15 & 1.09 & 1.05 & 1.02 & 0.95 & 0.93 & 5.31 & 4.09 & 3.11 & 2.31 & 1.70 & 1.27 & 0.96 & 0.74 & 0.59 & 0.49 & 0.42 & 0.37 & 0.34 & 0.28 & 0.26\\
&   100 & 5.83 & 4.62 & 3.66 & 2.89 & 2.31 & 1.91 & 1.62 & 1.43 & 1.30 & 1.21 & 1.16 & 1.12 & 1.09 & 1.04 & 1.03 & 5.26 & 4.05 & 3.08 & 2.31 & 1.72 & 1.30 & 1.00 & 0.79 & 0.65 & 0.55 & 0.48 & 0.44 & 0.41 & 0.36 & 0.36\\
&   300 & 5.79 & 4.58 & 3.63 & 2.89 & 2.33 & 1.94 & 1.67 & 1.49 & 1.37 & 1.28 & 1.23 & 1.20 & 1.17 & 1.13 & 1.13 & 5.22 & 4.01 & 3.05 & 2.30 & 1.74 & 1.34 & 1.05 & 0.85 & 0.71 & 0.62 & 0.56 & 0.52 & 0.50 & 0.46 & 0.45\\
&  $10^3$ & 5.76 & 4.56 & 3.62 & 2.89 & 2.35 & 1.98 & 1.72 & 1.55 & 1.43 & 1.36 & 1.31 & 1.28 & 1.26 & 1.23 & 1.23 & 5.19 & 3.98 & 3.03 & 2.30 & 1.75 & 1.37 & 1.10 & 0.91 & 0.79 & 0.70 & 0.65 & 0.61 & 0.59 & 0.56 & 0.55\\
 \hline
9&    12 & 5.91 & 4.71 & 3.72 & 2.91 & 2.28 & 1.84 & 1.53 & 1.32 & 1.17 & 1.07 & 0.99 & 0.94 & 0.90 & 0.80 & 0.77 & 5.35 & 4.14 & 3.15 & 2.32 & 1.69 & 1.24 & 0.91 & 0.68 & 0.51 & 0.40 & 0.32 & 0.26 & 0.22 & 0.13 & 0.11\\
&    30 & 5.91 & 4.70 & 3.72 & 2.91 & 2.30 & 1.87 & 1.56 & 1.36 & 1.21 & 1.12 & 1.05 & 1.01 & 0.98 & 0.90 & 0.88 & 5.34 & 4.13 & 3.14 & 2.33 & 1.70 & 1.26 & 0.94 & 0.72 & 0.56 & 0.45 & 0.38 & 0.33 & 0.29 & 0.23 & 0.21\\
&   100 & 5.88 & 4.67 & 3.69 & 2.91 & 2.31 & 1.89 & 1.60 & 1.40 & 1.27 & 1.18 & 1.12 & 1.08 & 1.05 & 1.00 & 0.99 & 5.31 & 4.10 & 3.11 & 2.32 & 1.72 & 1.29 & 0.98 & 0.76 & 0.62 & 0.52 & 0.45 & 0.40 & 0.37 & 0.32 & 0.31\\
&   300 & 5.81 & 4.61 & 3.65 & 2.89 & 2.32 & 1.93 & 1.65 & 1.46 & 1.33 & 1.25 & 1.20 & 1.16 & 1.14 & 1.09 & 1.09 & 5.25 & 4.03 & 3.07 & 2.31 & 1.73 & 1.32 & 1.03 & 0.82 & 0.68 & 0.59 & 0.52 & 0.48 & 0.46 & 0.42 & 0.41\\
&  $10^3$ & 5.76 & 4.56 & 3.62 & 2.89 & 2.35 & 1.97 & 1.71 & 1.54 & 1.42 & 1.34 & 1.29 & 1.26 & 1.24 & 1.21 & 1.20 & 5.20 & 3.99 & 3.04 & 2.30 & 1.75 & 1.36 & 1.09 & 0.90 & 0.77 & 0.68 & 0.62 & 0.59 & 0.56 & 0.53 & 0.52\\
 \hline
10&   360 & 5.85 & 4.64 & 3.68 & 2.93 & 2.36 & 1.96 & 1.68 & 1.49 & 1.37 & 1.28 & 1.23 & 1.19 & 1.17 & 1.13 & 1.12 & 5.28 & 4.07 & 3.10 & 2.34 & 1.76 & 1.35 & 1.06 & 0.86 & 0.72 & 0.62 & 0.56 & 0.52 & 0.49 & 0.45 & 0.44\\
   &400 & 5.83 & 4.63 & 3.67 & 2.92 & 2.36 & 1.97 & 1.70 & 1.51 & 1.39 & 1.30 & 1.25 & 1.22 & 1.19 & 1.15 & 1.15 & 5.27 & 4.06 & 3.09 & 2.34 & 1.77 & 1.36 & 1.07 & 0.87 & 0.73 & 0.64 & 0.58 & 0.54 & 0.51 & 0.47 & 0.47\\
   &600 & 5.81 & 4.61 & 3.66 & 2.92 & 2.37 & 1.99 & 1.72 & 1.54 & 1.42 & 1.34 & 1.29 & 1.26 & 1.23 & 1.20 & 1.19 & 5.24 & 4.03 & 3.08 & 2.33 & 1.78 & 1.38 & 1.10 & 0.90 & 0.77 & 0.68 & 0.62 & 0.58 & 0.56 & 0.52 & 0.52\\
 &$10^3$ & 5.80 & 4.60 & 3.65 & 2.92 & 2.38 & 2.00 & 1.73 & 1.55 & 1.43 & 1.36 & 1.31 & 1.27 & 1.25 & 1.22 & 1.21 & 5.24 & 4.03 & 3.07 & 2.33 & 1.78 & 1.39 & 1.11 & 0.92 & 0.78 & 0.70 & 0.64 & 0.60 & 0.58 & 0.54 & 0.54\\
 \hline
11&    10 & 5.40 & 4.19 & 3.25 & 2.53 & 2.02 & 1.67 & 1.43 & 1.28 & 1.19 & 1.13 & 1.09 & 1.07 & 1.06 & 1.04 & 1.04 & 4.84 & 3.62 & 2.67 & 1.94 & 1.42 & 1.06 & 0.81 & 0.65 & 0.54 & 0.48 & 0.44 & 0.41 & 0.40 & 0.38 & 0.38\\
&    30 & 5.28 & 4.10 & 3.20 & 2.53 & 2.05 & 1.73 & 1.51 & 1.37 & 1.28 & 1.23 & 1.20 & 1.18 & 1.16 & 1.15 & 1.15 & 4.71 & 3.52 & 2.61 & 1.93 & 1.45 & 1.11 & 0.89 & 0.74 & 0.64 & 0.58 & 0.54 & 0.52 & 0.51 & 0.49 & 0.49\\
&   100 & 5.43 & 4.23 & 3.29 & 2.60 & 2.12 & 1.80 & 1.60 & 1.47 & 1.39 & 1.34 & 1.31 & 1.29 & 1.28 & 1.27 & 1.27 & 4.86 & 3.65 & 2.71 & 2.01 & 1.52 & 1.19 & 0.98 & 0.84 & 0.75 & 0.70 & 0.66 & 0.64 & 0.63 & 0.62 & 0.62\\
 &  140 & 5.37 & 4.18 & 3.25 & 2.53 & 2.01 & 1.65 & 1.40 & 1.24 & 1.13 & 1.06 & 1.01 & 0.99 & 0.97 & 0.94 & 0.94 & 4.81 & 3.61 & 2.66 & 1.94 & 1.41 & 1.04 & 0.78 & 0.60 & 0.48 & 0.40 & 0.35 & 0.32 & 0.30 & 0.27 & 0.27\\
 &  180 & 5.37 & 4.17 & 3.24 & 2.53 & 2.02 & 1.66 & 1.42 & 1.26 & 1.16 & 1.09 & 1.05 & 1.02 & 1.00 & 0.98 & 0.98 & 4.80 & 3.60 & 2.66 & 1.94 & 1.42 & 1.05 & 0.80 & 0.62 & 0.51 & 0.43 & 0.38 & 0.35 & 0.33 & 0.31 & 0.31\\
 &  220 & 5.37 & 4.17 & 3.24 & 2.53 & 2.03 & 1.68 & 1.44 & 1.28 & 1.18 & 1.11 & 1.07 & 1.05 & 1.03 & 1.01 & 1.01 & 4.80 & 3.60 & 2.66 & 1.94 & 1.43 & 1.07 & 0.82 & 0.64 & 0.53 & 0.46 & 0.41 & 0.38 & 0.36 & 0.34 & 0.34\\
 &  260 & 5.37 & 4.17 & 3.24 & 2.54 & 2.03 & 1.69 & 1.45 & 1.30 & 1.20 & 1.13 & 1.09 & 1.07 & 1.05 & 1.03 & 1.03 & 4.81 & 3.60 & 2.66 & 1.95 & 1.43 & 1.07 & 0.83 & 0.66 & 0.55 & 0.48 & 0.43 & 0.40 & 0.38 & 0.36 & 0.36\\
 &  300 & 5.38 & 4.18 & 3.24 & 2.54 & 2.04 & 1.69 & 1.46 & 1.31 & 1.21 & 1.15 & 1.11 & 1.09 & 1.07 & 1.05 & 1.05 & 4.81 & 3.60 & 2.66 & 1.95 & 1.44 & 1.08 & 0.84 & 0.67 & 0.56 & 0.49 & 0.45 & 0.42 & 0.40 & 0.38 & 0.38\\
&  $10^3$ & 5.38 & 4.18 & 3.25 & 2.55 & 2.05 & 1.71 & 1.48 & 1.33 & 1.24 & 1.18 & 1.15 & 1.13 & 1.11 & 1.10 & 1.10 & 4.81 & 3.61 & 2.67 & 1.96 & 1.45 & 1.10 & 0.86 & 0.70 & 0.60 & 0.53 & 0.49 & 0.47 & 0.45 & 0.44 & 0.43\\
 \hline
12&    10 & 5.85 & 4.65 & 3.68 & 2.88 & 2.28 & 1.85 & 1.56 & 1.35 & 1.21 & 1.11 & 1.05 & 1.00 & 0.96 & 0.87 & 0.85 & 5.28 & 4.08 & 3.10 & 2.30 & 1.69 & 1.25 & 0.93 & 0.71 & 0.55 & 0.44 & 0.37 & 0.32 & 0.28 & 0.20 & 0.19\\
&    30 & 5.87 & 4.66 & 3.69 & 2.90 & 2.30 & 1.88 & 1.59 & 1.38 & 1.25 & 1.16 & 1.09 & 1.05 & 1.02 & 0.95 & 0.94 & 5.31 & 4.09 & 3.11 & 2.31 & 1.70 & 1.27 & 0.96 & 0.74 & 0.59 & 0.49 & 0.42 & 0.37 & 0.34 & 0.28 & 0.27\\
&   100 & 5.83 & 4.62 & 3.66 & 2.89 & 2.31 & 1.91 & 1.63 & 1.43 & 1.30 & 1.22 & 1.16 & 1.12 & 1.10 & 1.04 & 1.03 & 5.26 & 4.05 & 3.08 & 2.31 & 1.72 & 1.30 & 1.00 & 0.79 & 0.65 & 0.55 & 0.49 & 0.44 & 0.42 & 0.37 & 0.36\\
&   300 & 5.79 & 4.58 & 3.63 & 2.89 & 2.34 & 1.95 & 1.68 & 1.49 & 1.37 & 1.29 & 1.23 & 1.20 & 1.18 & 1.14 & 1.13 & 5.22 & 4.01 & 3.05 & 2.30 & 1.74 & 1.34 & 1.05 & 0.85 & 0.72 & 0.63 & 0.57 & 0.53 & 0.50 & 0.46 & 0.45\\
&  $10^3$ & 5.76 & 4.56 & 3.62 & 2.89 & 2.35 & 1.98 & 1.72 & 1.55 & 1.44 & 1.36 & 1.31 & 1.28 & 1.26 & 1.23 & 1.23 & 5.19 & 3.98 & 3.03 & 2.30 & 1.75 & 1.37 & 1.10 & 0.91 & 0.79 & 0.70 & 0.65 & 0.61 & 0.59 & 0.56 & 0.55\\
 \hline
13  &   200 & 5.88 & 4.67 & 3.71 & 2.95 & 2.38 & 1.98 & 1.70 & 1.51 & 1.38 & 1.30 & 1.24 & 1.21 & 1.18 & 1.14 & 1.13 & 5.31 & 4.10 & 3.13 & 2.37 & 1.79 & 1.37 & 1.08 & 0.87 & 0.73 & 0.64 & 0.57 & 0.53 & 0.50 & 0.46 & 0.45\\
    &   300 & 5.85 & 4.65 & 3.70 & 2.95 & 2.39 & 2.00 & 1.73 & 1.54 & 1.42 & 1.34 & 1.29 & 1.25 & 1.23 & 1.19 & 1.18 & 5.29 & 4.08 & 3.12 & 2.36 & 1.80 & 1.39 & 1.11 & 0.91 & 0.77 & 0.68 & 0.62 & 0.58 & 0.55 & 0.51 & 0.51\\
&  $10^3$ & 5.81 & 4.61 & 3.67 & 2.94 & 2.41 & 2.04 & 1.79 & 1.62 & 1.51 & 1.43 & 1.39 & 1.36 & 1.34 & 1.31 & 1.31 & 5.24 & 4.03 & 3.09 & 2.35 & 1.81 & 1.43 & 1.17 & 0.98 & 0.86 & 0.78 & 0.72 & 0.69 & 0.67 & 0.64 & 0.63\\
 \hline
14  &   200 & 5.92 & 4.71 & 3.75 & 2.98 & 2.40 & 2.00 & 1.72 & 1.52 & 1.39 & 1.31 & 1.25 & 1.22 & 1.19 & 1.14 & 1.13 & 5.35 & 4.14 & 3.17 & 2.40 & 1.81 & 1.39 & 1.10 & 0.89 & 0.74 & 0.65 & 0.58 & 0.54 & 0.51 & 0.46 & 0.46\\
 &   300 & 5.89 & 4.69 & 3.73 & 2.98 & 2.42 & 2.02 & 1.75 & 1.56 & 1.43 & 1.35 & 1.29 & 1.26 & 1.23 & 1.19 & 1.18 & 5.33 & 4.11 & 3.15 & 2.39 & 1.82 & 1.41 & 1.12 & 0.92 & 0.78 & 0.69 & 0.62 & 0.58 & 0.56 & 0.52 & 0.51\\
&  $10^3$ & 5.84 & 4.64 & 3.70 & 2.97 & 2.44 & 2.07 & 1.81 & 1.64 & 1.52 & 1.45 & 1.40 & 1.37 & 1.35 & 1.32 & 1.32 & 5.28 & 4.07 & 3.12 & 2.38 & 1.84 & 1.46 & 1.19 & 1.00 & 0.87 & 0.79 & 0.73 & 0.70 & 0.68 & 0.64 & 0.64\\
 \hline
15&   300 & 5.86 & 4.65 & 3.69 & 2.92 & 2.34 & 1.94 & 1.65 & 1.46 & 1.33 & 1.24 & 1.19 & 1.15 & 1.13 & 1.08 & 1.07 & 5.30 & 4.08 & 3.11 & 2.33 & 1.75 & 1.33 & 1.03 & 0.82 & 0.68 & 0.58 & 0.52 & 0.47 & 0.45 & 0.40 & 0.39\\
  &   600 & 5.82 & 4.61 & 3.66 & 2.91 & 2.36 & 1.97 & 1.70 & 1.51 & 1.39 & 1.31 & 1.25 & 1.22 & 1.20 & 1.16 & 1.15 & 5.25 & 4.04 & 3.08 & 2.33 & 1.76 & 1.36 & 1.07 & 0.87 & 0.74 & 0.65 & 0.59 & 0.55 & 0.52 & 0.48 & 0.48\\
&  $10^3$ & 5.79 & 4.59 & 3.65 & 2.91 & 2.37 & 1.99 & 1.73 & 1.55 & 1.43 & 1.35 & 1.30 & 1.27 & 1.25 & 1.22 & 1.21 & 5.23 & 4.02 & 3.06 & 2.32 & 1.77 & 1.38 & 1.10 & 0.91 & 0.78 & 0.69 & 0.63 & 0.60 & 0.57 & 0.54 & 0.53\\
 \hline
\end{tabular}
\caption{Values of $-\log_{10}\eta=-\log_{10}[\sum_i((f\cdot e_i)/(\sigma_i(e_i\cdot
e_i))^2]^{1/2}$ (where $\sigma_i$ is in units of $10^{-25}$s$^{-1}$) 
as a function of $\log_{10} \tau$ (with $\tau$ in seconds) and
$m_\chi$ in GeV, for each of the 15 decay channels whose numbering 
corresponds to the key in figs.\ \ref{fig1} or \ref{fig4}.  Figures on left for
WMAP7 and on right for Planck.  Values for longer lifetimes are the same as for $\log_{10} \tau=17$.}
\label{tab2}
\end{table*}

{\bf Acknowledgements.}  We thank Marco Cirelli, Tongyan Lin, Tracy Slatyer and Christoph Weniger for
very helpful clarifications and explanations, and
Jasper Hasenkamp for pointing out an inconsistency in earlier versions
of this work.  J.C.\ is supported by
the Natural Science and Engineering Research Council of Canada (NSERC).
P.S.\ is supported by the Banting program, administered by NSERC.

\bibliographystyle{apsrev}

\end{document}